\documentclass[
aip,
amsmath, amssymb,
reprint]{revtex4-1}
\usepackage[utf8]{inputenc}
\usepackage{amsmath}

\usepackage{float}
\usepackage{graphicx}
\usepackage[utf8]{inputenc}
\usepackage[T1]{fontenc}
\usepackage{mathptmx}
\usepackage{xcolor}

\begin{document}
\title{Dynamic Bandstructure and Capacitance Effects in Scanning Tunneling Spectroscopy of Bilayer Graphene}

\author{Gregory R. Holdman}\thanks{These authors contributed equally to this work.}
\affiliation{ 
University of Wisconsin-Madison, Department of Physics, 1150 University Ave., Madison, Wisconsin 53706, USA}
\author{Zachary J. Krebs}\thanks{These authors contributed equally to this work.}
\affiliation{ 
University of Wisconsin-Madison, Department of Physics, 1150 University Ave., Madison, Wisconsin 53706, USA}
\author{Wyatt A. Behn}\thanks{These authors contributed equally to this work.}
\affiliation{ 
University of Wisconsin-Madison, Department of Physics, 1150 University Ave., Madison, Wisconsin 53706, USA}
\author{Keenan J. Smith }
\affiliation{ 
University of Wisconsin-Madison, Department of Physics, 1150 University Ave., Madison, Wisconsin 53706, USA}
\author{K. Watanabe}
\affiliation{
National Institute for Materials Science, Namiki 1-1, Tsukuba, Ibaraki 305-0044, Japan}
\author{T. Taniguchi}
\affiliation{
National Institute for Materials Science, Namiki 1-1, Tsukuba, Ibaraki 305-0044, Japan}
\author{Victor W. Brar}\thanks{Author to whom correspondence should be addressed: vbrar@wisc.edu}
\affiliation{ 
University of Wisconsin-Madison, Department of Physics, 1150 University Ave., Madison, Wisconsin 53706, USA}

\date{\today}
       
\begin{abstract}
We develop a fully self-consistent model to describe scanning tunneling spectroscopy (STS) measurements of Bernal-stacked bilayer graphene (BLG), and we compare the results of our model to experimental measurements. Our results show that the STS tip acts as a top gate that changes the BLG bandstructure and Fermi level, while simultaneously probing the voltage-dependent tunneling density of states (TDOS). These effects lead to differences between the TDOS and the local density of states (LDOS); in particular, we show that the bandgap of the BLG appears larger than expected in STS measurements, that an additional feature appears in the TDOS that is an artifact of the STS measurement, and that asymmetric charge distribution effects between the individual graphene layers are observable via STS. 
\end{abstract}

\maketitle

Bilayer graphene (BLG) has been shown to display a rich electronic structure that is strongly dependent on both the electrostatic environment and the relative layer orientation. For example, transverse electric fields in Bernal-stacked BLG can induce a continuously tunable bandgap,\cite{zhang-direct-2009} while introducing a relative twist angle between the individual graphene sheets has been shown to promote correlated electron behavior such as superconductivity and Mott-like insulation. \cite{cao_correlated_2018,cao_unconventional_2018,yankowitz_tuning_2018,kerelsky_magic_2018,choi_imaging_2019} Furthermore, at certain temperatures and carrier densities, electrons in BLG exhibit hydrodynamic flow. \cite{Lucas-ElectronicHydro-2018,bandurin_negative_2016,crossno_observation_2015} These effects are, however, often sensitive to local perturbations which can alter the nature of novel electronic states, and obfuscate them in spatially averaged measurements; a complete understanding of BLG behavior, therefore, requires the development of local probes that can correlate electronic structure with crystal orientation, atomic defects, and charged impurities.

Scanning tunneling spectroscopy (STS) provides a promising pathway for understanding the role of disorder in BLG at atomic length scales. STS measurements of BLG on SiO$_2$ and hexagonal boron nitride (hBN) have already been used to probe a range of properties including quasiparticle dispersion,\cite{yankowitz-band-2014} gate-induced gap formation, \cite{Rutter2011,deshpande-mapping-2009,yankowitz-band-2014} localized bound or scattering states, \cite{Rutter2011,velasco2018visualization} and Landau level splittings that indicate correlated electron behavior. \cite{Rutter2011} The dynamic electronic structure of BLG, however, complicates the STS data. In particular, the BLG bandstructure is sensitive to electric fields applied perpendicular to the sample, which can both open a bandgap and change the carrier density. \cite{young-electronic-2012,zhang-direct-2009,jung_direct-2017,oostinga-gate-2008,delabarrera-theory-2015} In STS measurements, the perpendicular fields change as the tip voltage is varied, leading to differences between the local density of states (LDOS) and the tunneling density of states (TDOS). Understanding how to reconcile those differences in STS measurements is important when extracting physical BLG parameters such as bandgap, carrier density, or signatures of correlated electron behavior. 

In previous STS measurements, the effect of the changing electric field from the tip has been observed as instantaneous charging events where either a localized state \cite{velasco2018visualization} or Landau level \cite{Rutter2011} of the BLG was pushed across the Fermi level due to tip-induced doping.  In those results, and in other reports, \cite{deshpande-mapping-2009} the effect of tip induced changes in carrier density, and the effect of the back gate on the BLG band gap were considered. However, a full description requires a model that includes the tip's effect on the transverse field as well. In this study, we apply techniques from previously developed theoretical models of BLG in an electric field \cite{abergel-compressibility-2011,mccann-asymmetry-2006,young-capacitance-2011} to calculate how the electronic structure of the BLG changes with tip voltage, and how those changes affect the TDOS. We find that STS measurements of the BLG gap are expected to overestimate the width of the bandgap, and that an extra feature is introduced in the TDOS that is related to the BLG bandgap crossing the Fermi energy. These calculations also show how interlayer capacitance phenomena can be observed in STS measurements, and that -- due to offsetting effects -- the dependence of the charge neutral point on back gate voltage is not significantly altered by tip gating. We compare those calculations to experimental STS spectra taken from a BLG/hBN device at 4.5K and find good quantitative agreement.

\begin{figure*}
\includegraphics[width=16cm,keepaspectratio]{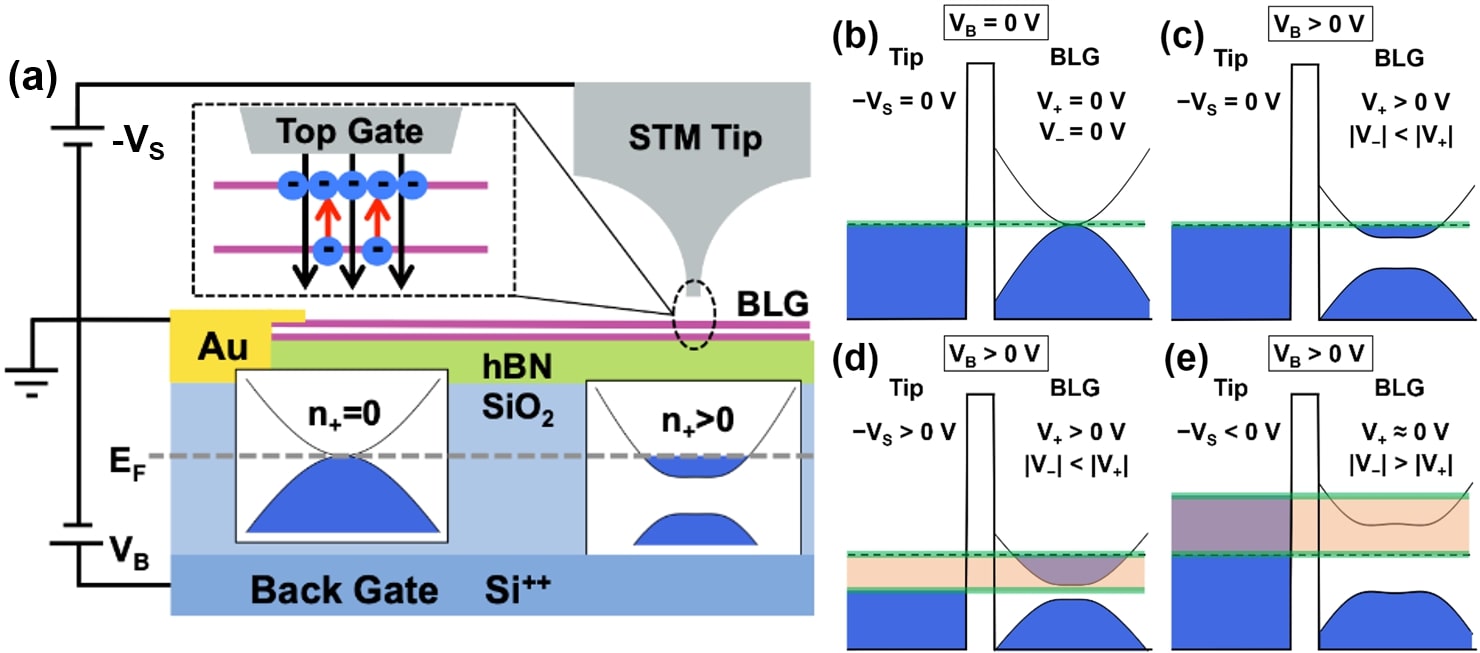}
\vspace{-3mm}
\caption{\label{fig:schematic} (a) Schematic of the STM setup with -V$_S\neq$ V$_B = 0$ V. A nonzero tunneling bias applied to the STM tip both locally dopes the sample and opens a bandgap. (b)-(e) Tunneling scenarios in selected electrostatic environments. Solid blue indicates filled electron states, orange shading indicates tunneling states, while green indicates states that are measured in a $dI/dV_S$ measurement at the given V$_S$.}
\vspace{-4mm}
\end{figure*}

A schematic of an STS measurement of BLG is shown in Fig. \ref{fig:schematic}(a). We assume the radius of curvature of the STM tip ($>$ 100 nm) is much larger than the tip-sample distance ($\lesssim$ 1 nm), so that the tip can be modeled as a flat plate. \cite{ChenSTM} In this geometry, the tip acts as a top gate with voltage set to $-V_S$ ($V_S$ is the sample bias) while a back gate voltage $V_B$ is also applied to the doped silicon under the hBN/SiO$_2$ layer. These gates induce a potential $V_1$ ($V_2$) on the top (bottom) layer of the BLG which may be combined into a symmetric and antisymmetric combination, $V_\pm=(V_1\pm V_2)/2$. Physically, $V_+$ represents the local electrostatic potential of the BLG patch under the STM tip, and $V_-$ is half of the electrostatic potential difference between the layers. Since the sample is grounded, the electrochemical potential is zero, $\Phi=\epsilon_F-eV_+$ = 0, with $\epsilon_F$ denoting the local Fermi level. For $\epsilon_F\neq 0$, the sample may be doped to a carrier density $n_+=n_1+n_2$, where $n_i$ is the carrier density on layer $i$. In general, $n_1\neq n_2$, and we denote the carrier layer asymmetry by $n_-=n_1-n_2$.

The bandstructure of BLG is dependent on the interlayer potential difference $V_-$, and a nonzero value of $V_-$ opens up a bandgap with magnitude \cite{mccann-asymmetry-2006}

\vspace{-3.0mm}
\begin{equation}
  \Delta_g = \frac{2\gamma_1e|V_-|}{\sqrt{\gamma_1^2+4e^2|V_-|^2}},
\end{equation}
where $\gamma_1=0.35$ eV is the interlayer hopping potential of BLG.\cite{mccann-electronic-2013,abergel-compressibility-2011} The relationship between $n_+$ and the Fermi level is also dependent on the interlayer potential energy $u=-2eV_-$. For low temperatures ($k_BT\ll \epsilon_F$), it can be shown that \cite{abergel-compressibility-2011} 
\vspace{-2.5mm}

\begin{equation}\label{FermiEqn}
\epsilon_F^2 = \frac{(\pi \hbar^2 v_F^2 n_+)^2 + \gamma_1^2u^2}{4(\gamma_1^2+u^2)}. 
\end{equation}

When the sample is undoped -- that is, $\epsilon_F$ lies within the bandgap -- the BLG is insulating and the system can be described by two gate electrodes filled with dielectric media. If the BLG is doped, however, the accumulated charge will establish equilibrium with the applied voltages to create an inter-sheet polarization field that partially offsets the applied field. An application of Gauss’ law gives the local potential of the sample just underneath the tip as
\begin{equation}
  V_+=\frac{d_1d_2}{\epsilon_0d_2+\epsilon_2d_1}\left(\frac{\epsilon_0}{d_1}(-V_S-V_-)+\frac{\epsilon_2}{d_2}(V_B+V_-)-en_+\right),
\end{equation}
and the bare, unscreened potential difference is given by
\begin{equation}\label{Charge_VM}
V_{-}^{ext} = \left( \frac{4\epsilon_0}{d} + \frac{\epsilon_0}{d_1} + \frac{\epsilon_2}{d_2} \right)^{-1}\left(\frac{\epsilon_0}{d_1}(-V_S-V_+) - \frac{\epsilon_2}{d_2}(V_B-V_+)\right).
\end{equation}
where $\epsilon_2$, $d_2$ are the dielectric constant and thickness of the substrate, $d_1$ is the tip height above the top BLG sheet, and $d$ is the intersheet spacing of the BLG. 

In our model we self-consistently solve the equation $ V_{-} = V_{-}^{ext}(V_+) + \frac{d_{eff}}{\epsilon_{eff}} en_{-} (V_+,V_-)$ for every pair of $V_{(S,B)}$, where $V_-^{ext}$ is the unscreened interlayer potential difference, and $\frac{d_{eff}}{\epsilon_{eff}}$ is the prefactor of Eq. \ref{Charge_VM}. This gives the carrier density and bandstructure of the BLG as a function of $V_S$, allowing us to compute the tunneling spectra $dI/dV_S$ (see Supplementary Online Materials). The free parameters used in our self-consistent model are the tip height $d_1$, the tip-sample work function difference $\Delta W_{t-s}$, and the back gate voltage offset due to substrate-induced doping $V_{B,0}$. We note that while it is justified to model the tip as a parallel plate in the electrostatic equations, $d_1$ is not necessarily equal to the true tunneling distance, since sharp protrusions may have a negligible contribution to the tip capacitance.

\begin{figure}
\label{fig:theoryresults:Figure2}
\includegraphics[width=8.5 cm, keepaspectratio]{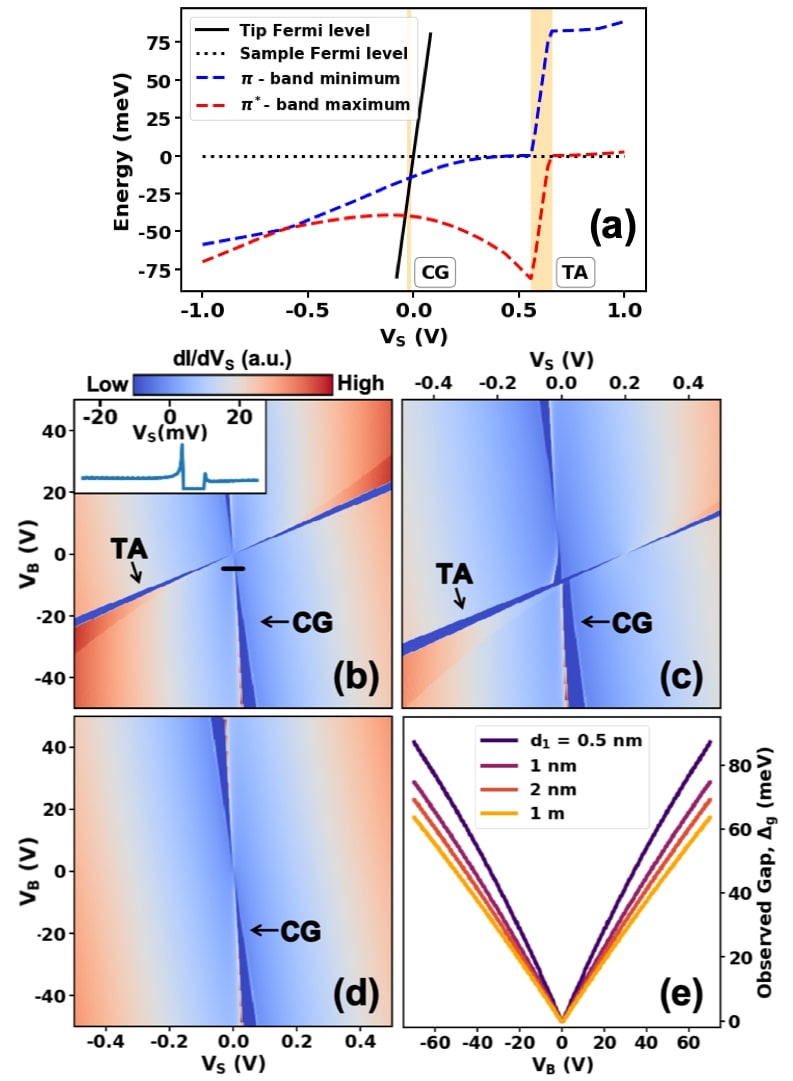}
\vspace{-2mm}
\caption{\label{fig:theoryresults} (a) Energies of BLG bandstructure features as a function of sample bias $V_S$ for $V_B=20$ V, $d_1=2$ nm, $d_2=285$ nm, and $\Delta W_{t-s}=0$. CG and TA occur within the indicated regions when the tip and sample Fermi levels, respectively, are aligned with the BLG bandgap. (b)-(d) Theoretical $dI/dV_S$ for the following model parameters: (b) $d_1=2$ nm, $d_2=285$ nm, $\Delta W_{t-s}=0$, $V_{B,0}=0$. Inset is a linecut along the black line at $V_B=-5$ V; (c) $d_1=2$ nm, $d_2=285$ nm, $\Delta W_{t-s}=-0.2$ eV, $V_{B,0}=0$; (d) $d_1=1$ m, $d_2=285$ nm, $\Delta W_{t-s}=0$ eV, $V_{B,0}=0$. (e) Width of the CG as a function of back gate voltage $V_B$ for different $d_1$. }
\vspace{-4mm}
\end{figure}

\begin{figure}[ht!]
\includegraphics[width=8.5 cm,keepaspectratio]{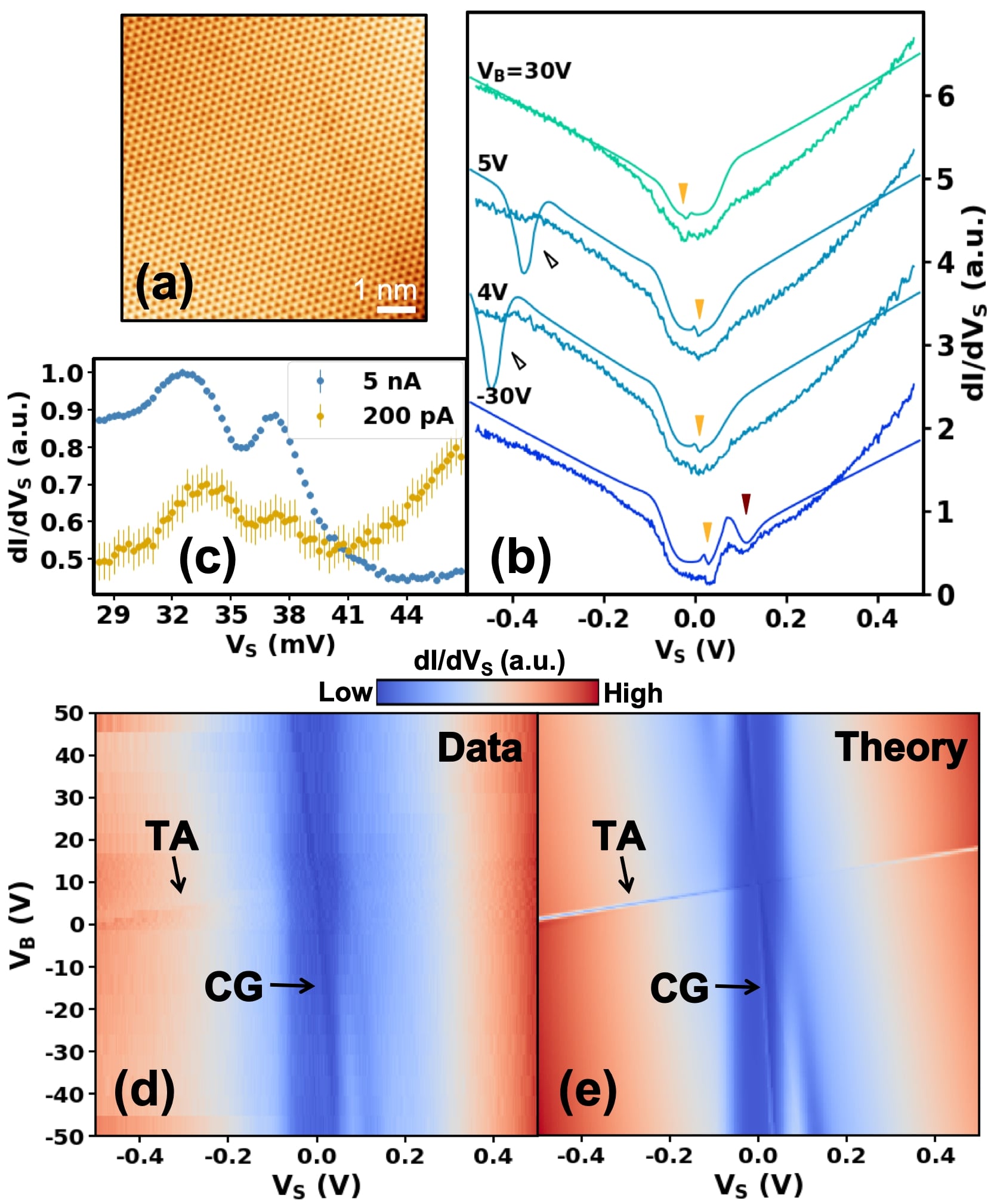}
\caption{\label{fig:experimental} (a) STM constant-current topographic image of BLG ($I=100$ pA, $V_S=-500$ mV). (b) Experimental (lower) and theoretical (upper) $dI/dV_S$ tunneling spectra at four back gate voltages: -30, +4,
+5, and +30 V. Curves are offset for clarity. Yellow (red) arrows indicate mid-gap point $E_D$ in the elastic (inelastic) signal. Hollow arrows point to TA. (c) $dI/dV_S$ measurements of the elastic CG as seen within the phonon gap at high (5 nA) and low (200 pA) tunneling current setpoints. ($V_B = 31$ V and V$_S$ setpoint is $-150$ mV; data taken from a different set of measurements) (d) Experimental $dI/dV_{S}$ gate map taken at a fixed location on the BLG flake (setpoint is $I = 3$ nA, $V_S = 700$ mV). Back gate resolution within the range V$_B$ = 0-20 V is 1 V, otherwise it is 5 V. (e) Simulated $dI/dV_S$ gate map fit to parameters that match data (see text). }
\vspace{-4mm}
\end{figure}

\begin{figure*}[t]
\includegraphics[width=17cm, keepaspectratio]{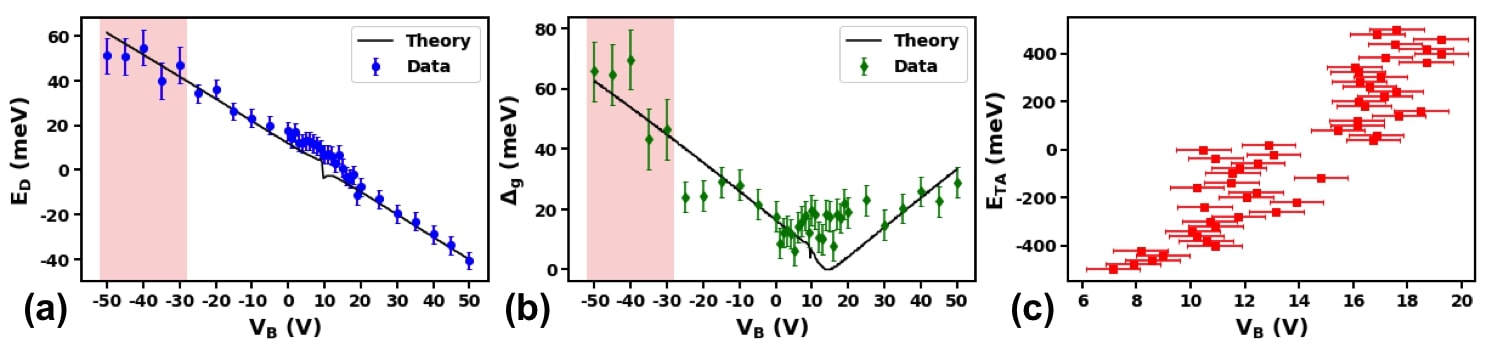}
\vspace{-3mm}
\caption{\label{fig:conv-gap-data} (a) The mig-gap points $E_D$ from Fig. \ref{fig:experimental}(d,e). Experimental values in the white (red) region are the minima in the elastic (inelastic) part of the spectrum. Inelastic values have had the 63 meV phonon energy subtracted. (b) CG width extracted from a fit to elastic (inelastic) tunneling spectra given in white (red) region. (c) Position of the TA, found by applying a piecewise fit function along $V_B$ axis.}
\vspace{-3mm}
\end{figure*}

Figures \ref{fig:schematic}(b-e) illustrate how the bandstructure of intrinsic BLG changes with varying V$_{(S,B)}$, and how those changes result in several key features in the TDOS. When the applied voltages are zero (Fig. \ref{fig:schematic}(b)), the BLG is charge neutral and the Fermi levels of the tip and sample align. For the case when $V_B > 0$ (Fig. \ref{fig:schematic}(c)), the BLG becomes n-doped and a bandgap opens due to a transverse electric field from the back gate. If a positive tip bias is also applied ($-V_S>0$) while keeping the same $V_B > 0$ (Fig. \ref{fig:schematic}(d)), an opposing electric field from the tip further dopes the BLG, and also partially closes the bandgap by reducing the transverse electric field. Changes in the tip voltage lead to changes in the states available for tunneling, yielding a nonzero $dI/dV_S$ signal. We identify contributions from both the tip and sample Fermi levels -- with the latter being caused by tip-induced band bending. If the tip Fermi level aligns with the BLG bandgap, no additional states are available at the lowest energies, which creates a minimum in the TDOS which we call the conventional gap (CG) feature. Meanwhile, a negative tip bias ($-V_S<0$, Fig. \ref{fig:schematic}(e)) will both widen the bandgap and decrease the carrier density of the BLG. In certain cases, the sample Fermi level can be pushed into the bandgap, again reducing the number of tunneling states and creating another minimum in the TDOS which we call the tunneling anomaly (TA).

For a fixed back gate voltage of V$_B = 20$ V, this behavior is shown quantitatively in Fig. \ref{fig:theoryresults}(a), where we plot the band minima/maxima as a function of sample bias. The CG is observed when the tip Fermi level aligns with the bandgap of BLG, while the TA is attributed to the sample Fermi level moving through the bandgap due to tip gating. We note that the underlying cause of the TA is similar to the mechanism of negative differential resistance (NDR) in other spectroscopic studies of atoms and molecules on insulating or semiconducting surfaces, as well as 2D materials. \cite{esaki1966new, tu2008controlling, grobis2005tuning, guisinger2005probing, repp2005molecules, lu2004molecular, zeng2000negative, lyo1989negative, gaudioso2000vibrationally, chen1999large, bedrossian1989demonstration} In those studies, the tip-induced band bending pushes the Fermi level through a molecular resonance, leading to charge-induced changes in the conductivity, while in this work the band bending pushes the BLG into an insulating state underneath the STM tip.

 The gate-dependent $dI/dV_S$ spectra calculated using our full self-consistent model are shown in Fig. \ref{fig:theoryresults}(b-d) for different choices of tip height and work function, and geometrical parameters consistent with a typical hBN/SiO$_2$ substrate. These results show that when fully accounting for the capacitive effects of the tip, the spectra exhibit several key features. First, the TA appears in the spectrum as a local minimum that disperses with V$_B$ having slope opposite to the CG. The TA is visible for small tip-sample distances (Fig. \ref{fig:theoryresults}(b,c)), but disappears when the tip is far from the sample (Fig. \ref{fig:theoryresults}(d)), demonstrating that this feature is due to the electrostatic influence of the tip. Second, the CG appears as a narrow dip that occurs between two sharp peaks that are due to the van Hove singularities (vHs) at the conduction and valence band edges of the BLG (inset Fig. \ref{fig:theoryresults}(b)). Our calculation shows asymmetry in the vHs peaks due to the broken layer symmetry in the presence of an electric field as well as preferential tunneling into the top layer of the BLG.\cite{mccann-asymmetry-2006} These predictions are consistent with previous STS measurements of Bernal stacked BLG, \cite{Rutter2011} as well as recent measurements performed on twisted bilayer samples. \cite{kerelsky_magic_2018,xie-spectroscopic-2019,choi_imaging_2019, jiang2019charge} Third, as shown in Fig. \ref{fig:theoryresults}(e), the apparent energy width of the CG is tip height dependent, such that for V$_B$ = 50 V, it appears as a 65, 55, and 51 meV gap for tip heights of 0.5, 1, and 2 nm, respectively. Fourth, there is an asymmetric increase in the $dI/dV_S$ signal on either side of the TA due to vHs on the band edge. As described above, this effect is stronger on one side due to the asymmetric occupation of charge on the layers, a phenomenon related to near-layer capacitance enhancement (NLCE), which has previously been observed in transport measurements. \cite{young-capacitance-2011,young-electronic-2012}

To compare our predictions to experimental measurements, we took STS measurements on exfoliated BLG samples at 4.5K and at $<10^{-11}$ mbar. Six separate experiments were performed on four BLG/hBN samples and two BLG/SiO$_2$ samples, all of which showed results consistent with the data shown here, with some variation attributed to microscopic changes in the STM tip and charge puddles in the SiO$_2$ (see Supplementary Online Materials). The thickness of the hBN layers in all devices were optically estimated to be around 100 nm (on 285 nm SiO$_2$) through comparison with samples measured by atomic force microscopy. Measurements were carried out with chemically etched STM tips made of Pt/Ir alloy. The differential conductance signal was obtained using a lock-in amplifier with a modulation amplitude of 0.25-7 mV and frequency of 200-700 Hz. Before taking data on the BLG, the spectroscopic integrity of the STM tip was verified by acquiring $dI/dV_S$ spectra on an Au(111) surface. \cite{chen1998scanning}

Figure \ref{fig:experimental}(a) shows a topographic image of the BLG surface. Point spectroscopy measurements using the same STM tip are given in Fig. \ref{fig:experimental}(b) for a few gate voltages, alongside simulated spectra calculated with our self-consistent model for $d_1=5.5$ nm, $W_t=4.45$ eV, and V$_{B,0}=-12$ V. The calculated spectra also include the effects of the known inelastic tunneling channel due to a phonon mode with energy $E_{\text{ph}} = 63$ meV (see Supplementary Online Materials).\cite{zhang-giant-2008} For gate voltages $-50 \leq$ V$_B$ $\leq$ 50 V, the experimentally measured and calculated $dI/dV_S$ spectra are shown side-by-side in Fig. \ref{fig:experimental}(d,e). These results show strong quantitative agreement. We observe a narrow CG within the elastic signal (the central dark blue region where $|V_S|<E_{\text{ph}}$) and a broadened inelastic CG signal that is shifted in energy by the phonon. Furthermore, we identify the TA as a narrow feature that has a strong dependence on the applied gate voltage and the NLCE asymmetry predicted in the model. A highly resolved spectroscopic measurement of the elastic CG is shown in Fig. \ref{fig:experimental}(c) for two different tunneling setpoints of 5 nA and 200 pA, representing a tip height change of 1.5 \AA. These data show two peaks of differing heights which we attribute to the vHs on the band edges. We measure a decrease in the peak spacing of $1\pm1$ meV as the tip retracts, which is consistent with Fig. 2(e), where we show that increases in tip height should decrease the bandgap. For a change in tip height of only 1.5 \AA, the change in observed bandgap is expected to be small for large (>2 nm) electrostatic distances, and is thus at the edge of our detection limit.

Unlike the CG, the energetic position of the TA is not shifted by inelastic tunneling, which is consistent with the predicted spectra. Within our model and fitting procedure, the CG and TA slopes are able to be simultaneously matched with the data, however, the theory predicts a more pronounced, gap-like TA. This discrepancy may be experimental in nature, due to thermal broadening in our 4.5 K STM, and broadening due to the lock-in modulation voltage. Another possibility is the breakdown of Eq. \ref{FermiEqn} near the TA, where the sample approaches charge-neutrality and one can no longer safely assume $k_BT \ll \epsilon_F$. Measurements of the TA obtained with smaller modulation voltages and different tip heights are shown in Fig. S6 in the Supplementary Online Materials.


We plot in Fig. \ref{fig:conv-gap-data} the CG and TA positions as a function of applied gate voltage, as well as the CG width. The CG position ($E_D$) and width ($\Delta_g$) were obtained by fitting points around the minimum in each spectrum to a piece-wise fit (see Supplementary Online Materials). The TA, meanwhile, appears as a discontinuity in our sample bias-dependent line cuts, and its position ($E_{TA}$) was determined based on the location of this discontinuity. These experimentally extracted parameters are compared to theoretical values from the results of Fig. \ref{fig:experimental}(e). Interestingly, the experimentally measured CG width $\Delta_g$ never drops to zero, hovering around 10-25 meV. This result is in disagreement with our model -- which predicts the CG gap to close regardless of the fitting parameters -- but is consistent with previous STS measurements of BLG.\cite{yankowitz-band-2014,Rutter2011} This discrepancy offers evidence for the appearance of gapped broken symmetry states when the applied electric field approaches zero, which are predicted to have an energy gap ranging from a few meV up to 30 meV. \cite{macdonald-gap, levitov-gap,yacoby-gap} However, it is also possible that this gap is caused by substrate interactions. In particular, the underlying hBN may apply a periodic potential to the BLG, which can create a persistent bandgap. \cite{ramasubramaniam-tunable-bandgaps-in-blg-bn-2011,mucha-heterostructures-2013,moon-electronic-2014} Finally, we note that our model uses a flat-plate capacitor model for the STM tip, while in reality it has some finite curvature which can create a localized potential well that confines the quasiparticles in the BLG such that a quantum dot is formed underneath the tip. \cite{velasco2018visualization, dombrowski1999tip} This quantization effect can also create persistent energy gaps in the STS spectrum.

In conclusion, we have demonstrated that a fully self-consistent electrostatic model for STS measurements of BLG is required to replicate many features of the $dI/dV_S$ spectrum. In particular, we show that the STM tip can act as a top gate which simultaneously modifies the carrier density and bandstructure of BLG. These effects are observable as a tunneling anomaly in the spectrum that has a gate-dependent slope that is opposite to that of the conventional band gap in BLG, and an overestimation of the BLG bandgap in TDOS measurements. Furthermore, the spectrum contains features related to the unique capacitative behavior of BLG, which can be understood through our electrostatic model. More generally, this work demonstrates the importance of considering tip-gating effects in STS experiments of 2D materials that are known to have electronic properties that depend on an applied perpendicular electric field, including twisted bilayer graphene \cite{lopes-graphene-2007,zou-band-2018} and some transition metal dichalcogenides. \cite{ramasubramaniam-tunable-2011,wu-electrical-2013,ross-electrical-2013} These results show that electrostatic models that include the effect of the varying tip voltage are necessary to relate the $dI/dV_S$ spectrum to the electronic structure of such materials.

See Supplementary Online Resources for additional information about our model and experiment.

Funding for this research was provided by NSF MRSEC under award number DMR-1839199. This research was performed using the compute resources and assistance of the UW-Madison Center For High Throughput Computing (CHTC) in the Department of Computer Sciences. The authors gratefully acknowledge use of facilities and instrumentation supported by NSF through the University of Wisconsin Materials Research Science and Engineering Center (DMR-1720415). K.W. and T.T. acknowledge support from the Elemental Strategy Initiative conducted by the MEXT, Japan, A3 Foresight by JSPS and the CREST (JPMJCR15F3), JST.

\bibliographystyle{aipnum4-1}
\bibliography{bibliography}{}

\end{document}